\documentclass{mpe_report}  
\usepackage{epsfig}  

\begin{document}  
\title{Compact star constraints on the high-density EoS}  
\author{H. Grigorian\inst{1,2,3}, D. Blaschke\inst{1,4,5}  
\and T. Kl\"ahn\inst{1,6}}  
\institute{ 
Institut f\"ur Physik, Universit\"at Rostock, 18051 Rostock, Germany  
\and  
Department of Physics, Yerevan State University, 375047 Yerevan, Armenia  
\and  
Laboratory for Information Technologies, JINR  
Dubna, 141980 Dubna, Russia 
\and  
Bogoliubov Laboratory for Theoretical Physics, JINR  
Dubna, 141980 Dubna, Russia 
\and 
Instytut Fizyki Teoretycznej, Uniwersytet Wroc{\l}awski, 
50-204 Wroc{\l}aw, Poland 
\and 
Gesellschaft f\"ur Schwerionenforschung mbH  (GSI), 
64291 Darmstadt, Germany 
} 
\maketitle  
  
\begin{abstract}  
A new scheme for testing the nuclear matter (NM) equation of state (EoS) 
at  high densities using constraints from compact star (CS) phenomenology  
is applied to neutron stars with a core of deconfined quark matter (QM). 
An acceptable EoS shall not to be in conflict with the mass measurement 
of 2.1 $\pm$ 0.2  M$_\odot$ (1 $\sigma$ level) for PSR J0751+1807 
and the mass-radius relation deduced 
from the thermal emission of RX J1856-3754.
Further constraints for the state of matter in CS 
interiors come from  temperature-age data for young, nearby objects.
The CS cooling theory shall agree not only with these data, but also 
with the mass distribution inferred via population  
synthesis models as well as with LogN-LogS data.
The scheme is applied to a set of hybrid EsoS with a phase transition to 
stiff, color superconducting QM which fulfills all above 
constraints and is constrained  otherwise from NM saturation 
properties and  flow data of heavy-ion collisions. 
We extrapolate our description to low temperatures and draw conclusions  
for the QCD phase diagram to be explored in heavy-ion collision experiments. 
\end{abstract}  
  
 \section{Introduction}  
Recently, new observational limits for the mass and the mass-radius  
relationship of CSs have been obtained which provide 
stringent constraints on the equation of state of strongly interacting  
matter at high densities, see \cite{Klahn:2006ir} and references therein. 
In this latter work several modern nuclear EsoS 
have been tested regarding their compatibility with phenomenology. 
It turned out that none of these nuclear EsoS meets 
all constraints whereas every constraint could have 
been fulfilled by some EsoS. 
As we will point out in this contribution, a phase transition 
to quark matter in the interior of CSs might resolve this problem. 
In the following we will apply an exemplary EoS for NM 
obtained from the ab-initio relativistic  
Dirac-Brueckner-Hartree-Fock (DBHF) approach using the Bonn A potential  
(\cite{DaFuFae05}). 
There is not yet an ab-initio approach to the high-density EoS formulated in 
quark and gluon degrees of freedom, since it  
would require an essentially nonperturbative treatment of QCD at finite  
chemical potentials. 
For some promising steps in the direction of a unified QM-NM description on the
quark level, we refer to the nonrelativistic potential model approach by
\cite{Ropke:1986qs} and the NJL model one by \cite{Lawley:2006ps}.
Simulations of QCD on the Lattice meet serious problems 
in the low-temperature finite-density domain of the QCD phase diagram relevant 
for CS studies. 
However, there are modern effective approaches to high-density QM  
which, albeit still simplified, focus on specific nonperturbative aspects of  
QCD.  
They differ from the traditional bag model approach and allow for CS  
configurations with sufficiently large masses, see \cite{Alford:2006vz}. 
For our QM description we employ a three-flavor chiral  
quark model of the NJL type with selfconsistent mean fields in the scalar 
meson (coupling $G_S$) and  scalar diquark (coupling $G_D=\eta_D~G_S$) 
channels (\cite{Blaschke:2005uj}), generalized by including  
a vector meson mean field (coupling $G_V=\eta_V~G_S$), see 
\cite{Klahn:2006iw}. 
  
We show that the presence of a QM core in the interior of CSs does not 
contradict any of the discussed constraints. 
Moreover, CSs with a QM interior would be assigned to the fast coolers 
in the CS temperature-age diagram. 
Another interesting outcome of our investigations 
is the prediction of a small latent heat for the deconfinement phase 
transition in both, symmetric and asymmetric NM.  
Such a behavior leads to hybrid stars that ``masquerade'' as neutron stars and 
has been discussed earlier by  \cite{Alford:2004pf} for a different EoS. 
This finding is of relevance for future heavy-ion collision programs at FAIR 
Darmstadt. 
 
\section{The flow constraint from HICs}  
 
The behaviour of elliptic flow in heavy-ion collisions is related  
to the EoS of isospin symmetric matter. 
The upper and lower limits  
for the stiffness deduced from such analyses  
(\cite{DaLaLy02}) 
are indicated in  Fig.~1 as a shaded region. 
\begin{figure}  
\centerline{
\epsfig{file=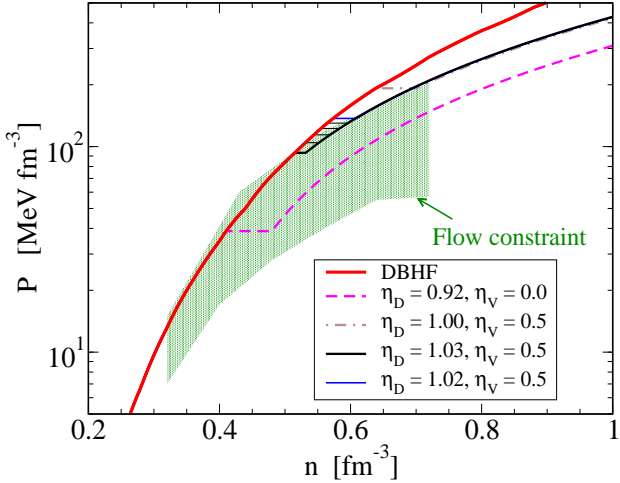,height=0.48\textwidth,width=0.38\textwidth,angle=-90}}  
\caption{Constraint on the high-density behavior of the EoS  
from simulations of flow data from heavy-ion collision experiments  
(shaded area from \cite{DaLaLy02}) compared to the nuclear matter and hybrid 
EsoS discussed in the text.  
\label{image5}}  
\end{figure}  
The nuclear DBHF EoS is soft at moderate densities with a compressibility  
$K=230$ MeV (\cite{DaFuFae04,boelting99}), but tends to violate  
the flow constraint for densities above 2-3 times nuclear saturation.  
As a possible solution to this problem we adopt a phase transition 
to QM with an EoS fixed to sketch the upper boundary of the flow constraint.  
In order to obtain an EoS as stiff as possible we use a vector coupling of  
$\eta_V=0.50$ and a diquark coupling of $\eta_D=1.03$. 
Herewith the EoS is completely fixed.  
 
\section{Constraints from astrophysics} 
  
\subsection{Maximum mass and mass-radius constraints}  
These most severe constraints come in particular from the mass  
measurement for PSR J0751+1807 (\cite{NiSp05}) giving a lower limit for 
the maximum mass $\approx 1.9~M_\odot$ at 1$\sigma$ level, 
and from the thermal emission of RX J1856-3754 (\cite{Trumper:2003we})
providing a lower limit in the  mass-radius plane with minimal radii 
$R>12$ km.  
These constraints can only be fulfilled by a rather stiff EoS.
The most stiff quark matter contribution to the EoS which still fulfills 
the flow constraint in symmetric matter corresponds to $\eta_V=0.5$ with
a maximum mass for hybrid stars $\approx 2.1~M_\odot$, rather independent
of the choice of $\eta_D$ which fixes the critical mass for
the onset of deconfinement, see Figs.~2,~3. 
For a more detailed discussion, see \cite{Klahn:2006ir,Klahn:2006iw}. 
 \begin{figure}  
\centerline{
\epsfig{file=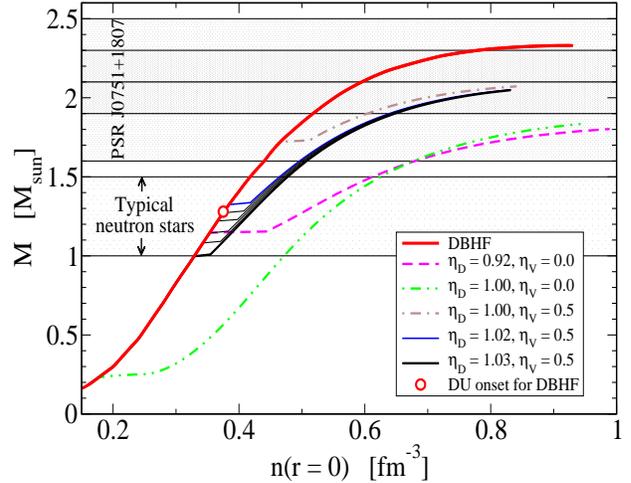,width=0.38\textwidth,height=0.48\textwidth,angle=-90} }  
\caption{ 
Stable CS configurations for neutron stars (DBHF) and hybrid stars, 
characterized by the parameters $\eta_D$ and $\eta_V$ of the quark matter EoS.
\label{image1}}  
\end{figure}   
\begin{figure}  
\centerline{
\epsfig{file=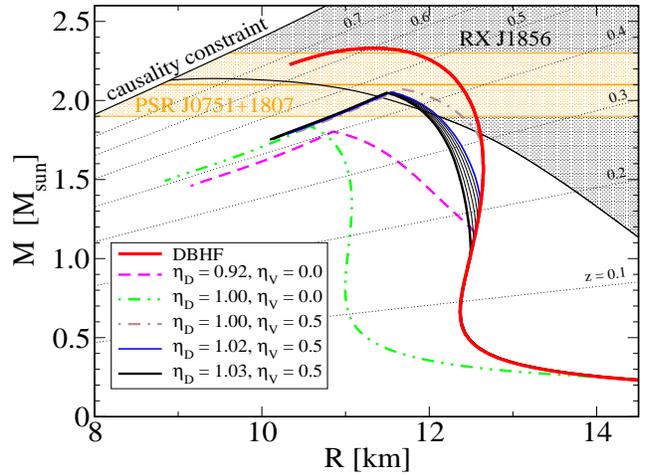,height=0.52\textwidth,width=0.38\textwidth,angle=-90} }  
\caption{Mass-radius relations for CSs with possible phase transition to  
deconfined quark matter, see \cite{Klahn:2006iw}.  
\label{image4}}  
\end{figure}

\subsection{Cooling constraints}  
  
{\it Direct Urca (DU) processes} are flavor-changing processes  
with the prototype being $ n\to p+e^-+\bar\nu_e$ (\cite{GaS41}), 
providing the most effective cooling mechanism in the hadronic layer of 
compact stars. It acts if  the proton fraction $x$ exceeds the 
DU threshold $x_{DU}$, $x=n_p/(n_n+n_p)\ge x_{DU}$. 
The threshold is given by $x_{DU}=0.11$ (\cite{Lattimer:1991ib}) 
and rises up to $x_{DU}=0.14$ upon inclusion of muons. 
Although the onset of the DU process entails  
a sensible dependence of cooling curves on the  
star masses, hadronic cooling with realistic pairing gaps is not 
sufficient to explain young, nearby X-ray dim objects,  
like Vela, with typical CS masses,  
not exceeding $1.5 ~M_\odot$ (\cite{Blaschke:2000dy,Grigorian:2004jq}). 
The point on the stability curve in Fig.~2 marks 
the DU threshold density for the DBHF EoS. 
Quark matter DU processes provide enhanced cooling, 
characterized by the diquark pairing gaps (\cite{Blaschke:1999qx,Page:2000wt}) 
and their density dependence (\cite{Grigorian:2004jq,Popov:2005xa}). 
For a recent review, see  \cite{Sedrakian:2006mq}.

To verify this  rather heuristic approach we apply 
explicit calculations of the cooling of hybrid configurations 
which shall describe present data of the  
{\it temperature-age} distribution of CSs. 
The main processes in nuclear matter that we 
accounted for are the direct Urca, the medium modified Urca and 
the pair breaking and formation processes. 
Furthermore we accounted for the $1S_0$ neutron 
and proton gaps and the suppression of the $3P_2$ 
neutron gap.  
For the calculation of the cooling of the 
quark core we incorporated the most efficient 
processes, namely the quark modified Urca process,  
the quark bremsstrahlung, the electron bremsstrahlung 
and the massive gluon-photon decay. 
In the 2-flavor superconducting phase 
one color of quarks remains unpaired. 
Here we assume a small residual pairing ($\Delta_X$) of the 
hitherto unpaired quarks. 
For detailed discussions of cooling 
calculations and the required ingredients see  
\cite{Blaschke:2000dy,Popov:2005xa,Blaschke:2006tt}
and references therein. 
The resulting temperature-age relations  
for the introduced hybrid EoS are shown in Fig.~4.  
The critical density for 
the transition from nuclear to quark matter has been 
set to a corresponding CS mass of $M_{\rm crit}=1.22~M_\odot$. 
All cooling data points are covered and correspond to CS configurations 
with reasonable masses. 
In this picture slow coolers correspond to light, 
pure neutron stars ($M<M_{\rm crit}$), whereas fast coolers are rather 
massive CSs ($M>M_{\rm crit}$) with a QM core.  
\begin{figure}  
\centerline{
\epsfig{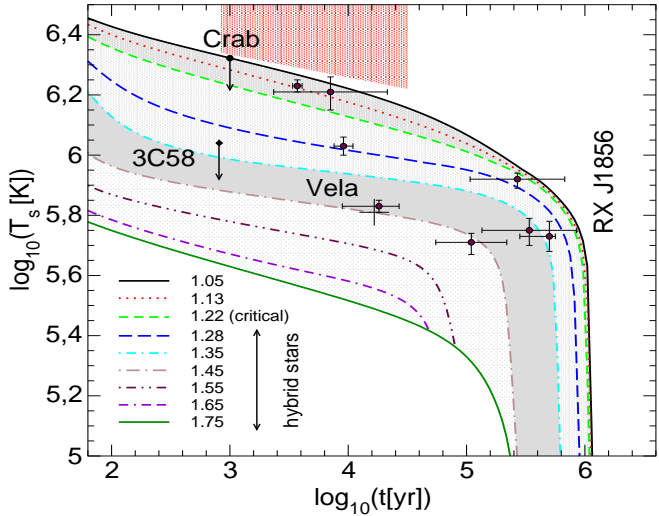} }  
\caption{Cooling evolution for hybrid stars of different masses given in  
units of $M_\odot$.  
Note that Vela is described with a typical CS mass not exceeding  
1.45 $M_\odot$. From \cite{Popov:2005xa}.   
\label{image2}}  
\end{figure}  
 
Another constraint on the temperature-age relation is given by the 
{\it maximum brightness} of CSs, as discussed by 
\cite{Grigorian:2005fd}. 
It is based on the fact that despite many observational efforts 
one has not observed very hot NSs ($\log$T $> 6.3-6.4$ K) 
with ages of $10^3$ - $10^{4.5}$ years. 
Since it would be very easy to find them - if they exist in the galaxy - 
one has to conclude that at least their fraction is very small. 
Therefore a realistic model  should not predict CSs with typical masses 
at temperatures noticeable higher than the observed ones. 
The region of avoidance is the hatched trapezoidal region in Fig. 4. 
 
The final CS cooling constraint in our scheme is given by the 
{\it Log N--Log S} distribution, 
where $N$ is the number of sources with observed fluxes larger than $S$. 
This integral distribution grows towards lower fluxes and is inferred, e.g.,
from the ROSAT all-sky survey (\cite{Truemper:1998zq}). 
The observed {\it Log N--Log S} distribution 
is compared with the ones 
calculated in the framework of a population synthesis approach in  Fig.~5.
A detailed discussion of merits and drawbacks can be found in 
\cite{Popov:2004ey}.
    
Altogether, the hybrid star cooling behavior  
obtained for our EoS fits all of the sketched constraints  
under the assumption of the existence of a 2SC phase  with X-gaps. 
  
\begin{figure}  
\centerline{
\epsfig{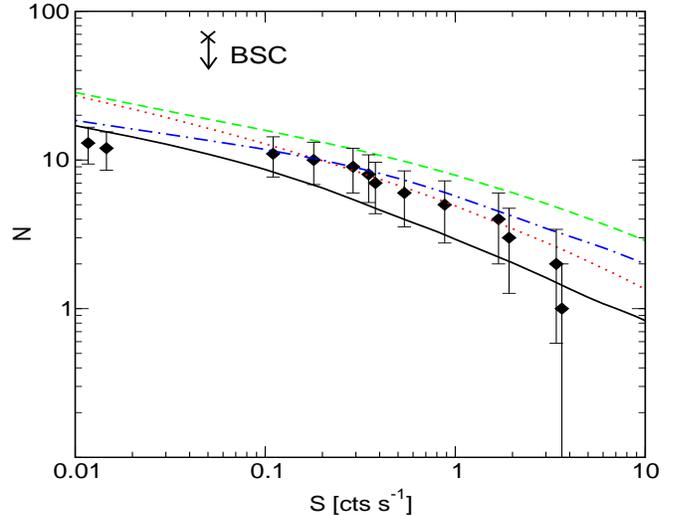} }  
\caption{Comparison of observational data for the LogN-LogS distribution 
with results from population synthesis using hybrid star cooling  
according to \cite{Popov:2005xa}.  
\label{image3}}  
\end{figure}  
  
\section{Outlook: The QCD phase-diagram}  
Within the previous sections we exemplified how 
to apply the testing scheme introduced in \cite{Klahn:2006ir} 
to the modeling of a reliable hybrid EoS with a NM-QM phase transition 
that fulfills a wide range of constraints from HICs and astrophysics. 
In a next step we extend the description 
to finite temperatures focusing on 
the behaviour at the transition line. 
For this purpose we apply  a relativistic mean-field model with 
density-dependent masses and couplings (\cite{Typel:2005ba})
adapted such as to mimick the DBHF-EoS and generalize to finite temperatures
(DD-F4).
Fig.~6 shows the resulting phase diagram including  
the transition from nuclear to quark matter  ($\eta_D=1.030$, $\eta_V=0.50$)  
which exhibits almost a crossover transition with a negligibly small 
coexistence region and a tiny density jump. 
At temperatures beyond $T\sim 45$ MeV our NM description is not reliable any 
more since contributions from mesons, hyperons and nuclear resonances 
are missing. 
This will be amended in future studies. 
 
\begin{figure}  
\centerline{
\epsfig{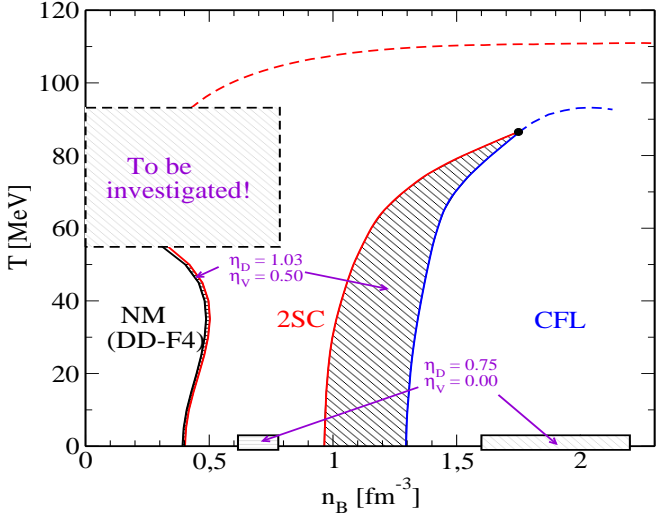}} 
\caption{Phase diagram for isospin symmetry using the most favorable  
hybrid EoS of the present study.  
The NM-2SC phase transition is almost a crossover.  
The model DD-F4 is used as a finite-temperature  
extension of DBHF. For the parameter set ($\eta_D=0.75$, $\eta_V=0.0$) 
the flow constraint is fulfilled but no stable hybrid stars are obtained. 
\label{image6}}  
\end{figure}  
  
\section{Conclusions}  
  
We have presented a new scheme for testing nuclear matter equations of state  
at supernuclear densities using constraints from neutron star and HIC
phenomenology.  
Modern constraints from the mass and mass-radius-relation measurements  
require stiff EsoS at high densities, whereas flow data from heavy-ion  
collisions seem to disfavor too stiff behavior of the EoS. 
As a compromise we have presented a hybrid EoS with a phase transition to  
color superconducting quark matter which, due to a vector meson meanfield, 
is stiff enough at high densities to allow compact stars with a mass 
of 2 $M_\odot$.  
Such a hybrid EoS could be superior to a purely hadronic one as it allows a  
faster cooling of objects within the typical CS mass region. 
This way, young nearby X-ray dim objects such as Vela could be explained  
with masses not exceeding 1.5 $M_\odot$.   
The present hybrid EoS predicts hybrid stars that ``masquerade'' as neutron  
stars, suggesting only a tiny density jump at the phase transition. 
This characteristics is also present for the symmetric matter case  
and persists at higher temperatures in the QCD phase diagram. 
It is suggested that the CBM experiment at FAIR might softly enter  
the quark matter domain without extraordinary hydrodynamical effects from 
the deconfinement transition. 
 
\vskip 0.4cm  
  
\begin{acknowledgements}  
We thank all our collaborators who have contributed to these results,  
in particular D. Aguilera, J. Berdermann, C. Fuchs, S. Popov,  
F. Sandin, S. Typel,  and D.N. Voskresensky.  
  The work is supported by DFG under grant 436 ARM 17/4/05 and by the 
Virtual Institute VH-VI-041 of the  Helmholtz Association.  
We also gratefully acknowledge the support by J.~E.~Tr\"umper and the  
organizers of the 363$^{\rm rd}$ Heraeus seminar on "Neutron Stars and  
Pulsars".  
\end{acknowledgements}  
  

\end{document}